\documentclass[10pt,journal,compsoc]{IEEEtran}
\ifCLASSINFOpdf
\else
\fi
\usepackage{graphicx}
\usepackage{algorithm,algorithmic}
\usepackage{amsmath}
\usepackage{amsfonts}
\usepackage{multirow}
\usepackage{wasysym}
\usepackage{color} 
\usepackage{soul}
\usepackage{listings}
\usepackage{lipsum}
\usepackage{mathtools}
\usepackage{cuted}

\begin{document}
%
\title{iBrownout: An Integrated Approach for Managing Energy and Brownout in Container-based Clouds}
%
%
%

\author{Minxian~Xu,~\IEEEmembership{}
	Adel Nadjaran Toosi,~\IEEEmembership{Member, ~IEEE,}
	and~Rajkumar Buyya,~\IEEEmembership{Fellow,~IEEE}
	\IEEEcompsocitemizethanks{\IEEEcompsocthanksitem M. Xu, A. N. Toosi and R. Buyya are with Cloud Computing and Distributed Systems (CLOUDS) lab, School of Computing and Information Systems,
		University of Melbourne, Australia, 3010
		.\protect\\
	}
	\thanks{Manuscript received： ; revised }}

%
%
\newcommand\MYhyperrefoptions{bookmarks=true,bookmarksnumbered=true,
	pdfpagemode={UseOutlines},plainpages=false,pdfpagelabels=true,
	colorlinks=true,linkcolor={black},citecolor={black},urlcolor={black},
	pdftitle={Bare Demo of IEEEtran.cls for Computer Society Journals},
	pdfsubject={Typesetting},
	pdfauthor={Michael D. Shell},
	pdfkeywords={Computer Society, IEEEtran, journal, LaTeX, paper,
		template}}

\markboth{Journal of \LaTeX\ Class Files,~Vol.~6, No.~1, January~2007}%
{Shell \MakeLowercase{\textit{et al.}}: Bare Demo of IEEEtran.cls for Journals}
%




\IEEEtitleabstractindextext{%
\begin{abstract}
Energy consumption of Cloud data centers has been a major concern of \color{black}many \color{black} researchers, and one of the reasons for huge energy consumption of Clouds lies in the inefficient utilization of computing resources. 
Besides energy consumption, another challenge of data centers is the unexpected loads, which leads to the overloads and performance degradation. Compared with VM consolidation and Dynamic Voltage Frequency Scaling that cannot function well when the whole data center is overloaded, brownout has shown to be a promising technique to handle both overloads and energy consumption through dynamically deactivating application optional components, which are also identified as containers/microservices. 
In this work, we propose an integrated approach to manage energy consumption and brownout in container-based cloud data centers. \color{black} We also evaluate our proposed scheduling policies with real traces in a prototype system. \color{black} \color{black} The results show that our approach reduces about 40\%, 20\% and 10\% energy than the approach without power-saving techniques, brownout-overbooking approach and auto-scaling approach respectively while ensuring Quality of Service.   \color{black}
\end{abstract}


}
\maketitle

\IEEEdisplaynontitleabstractindextext

%
\IEEEpeerreviewmaketitle

\section{Introduction}
Cloud computing has been regarded \color{black}as \color{black} a new paradigm for resource and service \color{black} provisioning\color{black}, which provides the pay-as-you-go pricing model \cite{Buyya}. Clouds have offered vital benefits for IT industry by \color{black}relieving the need for \color{black} building own infrastructures, therefore, the companies are able to concentrate on making profits with their services. In addition, innovative ideas and Internet technologies can also be delivered with less hardware investment and human expense. To support the proliferation of cloud services, more data centers are established, and many cloud service providers, like Google, Amazon and Microsoft are deploying their data centers around the world and offering their services. 

Although cloud data centers are providing compelling features for customers, the energy consumption of data centers has become a major \color{black}topic of research. \color{black} U.S. data centers have consumed 100 billion kWh electricity in 2015, which is equivalent to the total energy consumption of Washington City. It is estimated that the energy consumption of U.S. data centers will continue increasing and reach 140 billion kWh by 2020 \cite{Tom}\cite{Pierre}. The servers hosted in data centers dissipate heat and need to be maintained by cooling infrastructure, which provides the cooling resource to extract the heat from IT devices. Though the cooling infrastructure is already efficient \color{black}to some extent, \color{black} the servers are still one of the major energy consumers. Cloud data centers not only consume huge energy consumption, but also have a non-negligible impact on the environment. It is reported that data \color{black} centers have contributed 200 million metric tons of carbon dioxide to the environment \cite{Carbon2015}. \color{black} Recently, some dominant service providers \color{black} established \color{black} a community to promote energy efficiency for data centers to minimize the impact on the environment, which is also known as Green Grid \color{black} \cite{Beloglazov}. \color{black}  

However, reducing energy consumption is a challenging objective as applications and data are growing fast and complex \cite{liu2012renewable}. Normally, the applications and data are required to be processed within the required time, thus, large and \color{black}powerful \color{black} servers are required to offer services. To ensure the sustainability of future growth of data centers, cloud data centers \color{black}must be \color{black}designed to be efficiently utilize the \color{black} resources of \color{black} infrastructure and minimize energy consumption. To address this problem, the concept of green cloud is proposed, which aims to manage cloud data centers in an energy efficient manner \cite{Beloglazov}.  \color{black} Consequently, \color{black} data centers are required to offer resources while satisfying Quality of Service (QoS), as well as reduce energy consumption.

One of the main reasons of high energy consumption \color{black} of cloud data \color{black} centers lies in that computing resources are inefficiently utilized by applications on servers. Therefore, applications are currently built with microservices to utilize infrastructure resource more efficiently. Microservices are also referred as a set of self-contained application components \cite{Newman}. The components encapsulate its logic and expose its functionality via interfaces to make them flexible to be deployed and replaced. With microservices or components, developers and user benefit from their technological heterogeneity, resilience, scalability, ease of deployment, organizational alignment, composability and optimization for replicability. It also brings the advantage of more fine-grained control over the application resource usage. 

Thus, \color{black} in this paper, \color{black} we take advantage of \textbf{brownout}, a paradigm inspired from voltage shutdown that copes with emergency cases. In original brownout scenario, the light bulbs emit fewer lights to save energy consumption.  In Cloud scenario, brownout can be applied to microservices or application components that are allowed \color{black} to be temporarily deactivated without affecting main functionality. \color{black} When brownout is triggered, the user's experience is temporally degraded to \color{black} relieve the overloaded situation and reduce energy consumption. \color{black} 

It is common for microservices or application components to have this brownout feature. Klein et al. \cite{Klein} introduced an online shopping system that has a recommendation engine to recommend products to users. The recommendation engine enhances the function of the whole system, while it is not necessary to keep it running all the time, especially under the overloaded situation. As the recommendation engine requires more resource in comparison to other components, if it is deactivated, more clients with essential requests or QoS constraints can be served. Apart from this example, brownout paradigm is also suitable for other systems that allow \color{black} some \color{black} microservices or application components to not keep running all the time.

In this paper, we propose a brownout prototype system based on containers to reduce data center energy while ensuring Quality of Service. The main \textbf{contributions} of our work are as \color{black} follows: 
1)	Proposed an effective architecture that enables brownout paradigm to manage the container-based environment, which enables fine-grained control on containers;
2)	Presented several scheduling policies for managing microservices or containers to achieve power saving and QoS constraints;
3)	Implemented a prototype system and 4) carried out the evaluation in INRIA Grid'5000 testbed using resources from Lyon cluster for Wikipedia web workload. \color{black} 

The rest of this paper is organized as: Section 2 discusses the related work, followed by scenarios that brownout can be applied and the challenges for using brownout presented in Section 3. Section 4 and Section 5 introduce the architecture that enables brownout to manage the microservices or application components and models respectively. Scheduling policies for determining the activation and deactivation of microservices are presented in Section 6. In Section 7, we present our experiments environment and evaluate the performance of different scheduling policies. Conclusions and future directions are given in Section 8.

\section{Related Work}

\begin{table*}[]
	\centering
	\caption{Comparison of focus of related work and our work}
	\resizebox{1.0\textwidth}{!}{%
		\label{my-label}
		\begin{tabular}{|c|c|c|c|c|c|c|c|c|c|c|}
			\hline
			\multirow{2}{*}{\textbf{Approach}} & \multicolumn{3}{c|}{\textbf{Technique}} & \multicolumn{3}{c|}{\textbf{Optimization Objective}} & \multicolumn{2}{c|}{\textbf{Management Unit}} & \multicolumn{2}{c|}{\textbf{Experiments Platform}} \\ \cline{2-11} 
			& \textbf{\begin{tabular}[c]{@{}c@{}}VM \\ Consolidation\end{tabular}} & \textbf{DVFS} & \textbf{Brownout} & \textbf{\begin{tabular}[c]{@{}c@{}}Energy \\ Consumption\end{tabular}} & \textbf{SLA/QoS} & \textbf{Overloads} & \textbf{VMs} & \textbf{Containers} & \textbf{Simulation} & \textbf{ \color{black} Real Testbed} \\ \hline
			Beloglazov et al. \cite{AntonTPDS} & $\checkmark$ & $\times$ & $\times$ & $\checkmark$  & $\checkmark$  & $\checkmark$ & $\checkmark$  & $\times$ & $\checkmark$  & $\times$ \\ \hline
			Beloglazov et al. \cite{Beloglazov2} & $\checkmark$  & $\times$ & $\times$ & $\checkmark$  & $\checkmark$  & $\times$ & $\checkmark$  & $\times$ & $\checkmark$  & $\times$ \\ \hline
			Chen et al. \cite{Chen} & $\checkmark$  & $\times$ & $\times$ & $\checkmark$  & $\checkmark$  & $\times$ & $\checkmark$  & $\times$ & $\checkmark$  & $\times$ \\ \hline
			Han et al. \cite{Han} & $\checkmark$  & $\times$ & $\times$ & $\checkmark$  & $\checkmark$  & $\checkmark$  & $\checkmark$  & $\times$ & $\checkmark$  & $\times$ \\ \hline
			Mastroianni et al. \cite{Mastroianni} & $\checkmark$  & $\times$ & $\times$ & $\checkmark$  & $\times$ & $\checkmark$ & $\checkmark$ & $\times$ & $\times$ & $\checkmark$  \\ \hline
			Zheng et al. \cite{Zheng} & $\checkmark$  & $\times$ & $\times$ & $\checkmark$  & $\checkmark$  & $\times$ & $\checkmark$  & $\times$ & $\times$ & $\checkmark$  \\ \hline
			Ferdaus et al. \cite{Ferdaus} & $\checkmark$  & $\times$ & $\times$ & $\checkmark$  & $\checkmark$  & $\times$ & $\checkmark$  & $\times$ & $\times$ & $\checkmark$  \\ \hline
			Kim et al. \cite{Kim} & $\times$ & $\checkmark$  & $\times$ & $\checkmark$  & $\checkmark$  & $\times$ & $\times$ & $\times$ & $\checkmark$  & $\times$ \\ \hline
			Pietri et al. \cite{Pietri} & $\times$ & $\checkmark$  & $\times$ & $\checkmark$  & $\checkmark$  & $\times$ & $\times$ & $\times$ & $\checkmark$  & $\times$ \\ \hline
			Teng et al. \cite{Teng} & $\checkmark$  & $\checkmark$  & $\times$ & $\checkmark$  & $\checkmark$  & $\times$ & $\times$ & $\times$ & $\checkmark$  & $\checkmark$  \\ \hline
			Klein et al. \cite{Klein} & $\times$ & $\times$ & $\checkmark$  & $\times$ & $\checkmark$  & $\checkmark$  & $\times$ & $\times$ & $\times$ & $\checkmark$  \\ \hline
		    Tomas et al. \cite{tomas} & $\times$ & $\times$ & $\checkmark$  & $\times$ & $\checkmark$  & $\checkmark$  & $\times$ & $\times$ & $\times$ & $\checkmark$  \\ \hline
			Wang et al. \cite{Wang} & $\checkmark$  & $\times$ & $\times$ & $\checkmark$  & $\checkmark$  & $\times$ & $\checkmark$  & $\times$ & $\times$ & $\checkmark$  \\ \hline
			Xu et al. \cite{Xu} & $\checkmark$  & $\times$ & $\checkmark$  & $\checkmark$  & $\times$ & $\checkmark$  & $\checkmark$  & $\checkmark$  & $\checkmark$  & $\times$ \\ \hline
			Xu et al. \cite{xu2017energy} & $\checkmark$  & $\times$ & $\checkmark$  & $\checkmark$  & $\times$ & $\checkmark$  & $\checkmark$  & $\checkmark$  & $\checkmark$  & $\times$ \\ \hline
			iBrownout & $\times$ & $\times$ & $\checkmark$  & $\checkmark$  & $\checkmark$  & $\checkmark$  & $\times$ & $\checkmark$  & $\times$ & $\checkmark$  \\ \hline
		\end{tabular}
	}
\end{table*}

A recent report suggests that U.S. data center will consume 140 billion kWh of electricity annually in the next four years by 2020 \color{black} \cite{Tom}\color{black},  which equals to the annual output of about 50 brown power plants and \color{black} translates \color{black} to higher carbon emissions.  To decrease operational costs and environmental impact, numerous state-of-the-art research has been conducted to reduce data center energy consumption. The main categories for handling this energy efficient problem are VM consolidation and Dynamic Voltage Frequency Scaling (DVFS). 

VM consolidation minimizes energy consumption by allocating tasks among fewer machines and turning the unused machines into low-power mode or power-off state. To reduce the number of \color{black} active \color{black} machines, the VMs hosted on underutilized machines are consolidated to other machines and the underutilized machines are transformed into low-power mode. Beloglazov et al. \cite{AntonTPDS} proposed several VM consolidation algorithms to save data center energy consumption. The VM consolidation process is modeled as a bin-packing problem, where VMs are regarded as items and hosts are regarded as bins. The objective of these VM consolidation algorithms is mapping the VMs to hosts in an energy-efficient manner.  This work \color{black} advanced \color{black} the existing work by modeling the algorithms to be independent of workload types and do not need to know the VM application information in advance.  However, the algorithms have not been evaluated under realistic testbeds. Based on the VM consolidation approaches in this work, other works like \cite{Beloglazov2}\cite{Chen}\cite{Han}, have done some extension work to improve algorithm performance.

Mastroianni et al. \cite{Mastroianni} introduced a self-adaptive method for VM consolidation on both CPU and memory. The method aims to reduce the overall costs caused by energy-related issues. The VM consolidation process is determined by a probabilistic function based on Bernoulli trial. Both the mathematical analysis and realistic testbed results show that the proposed method reduces total energy consumption efficiently.   

Zheng et al. \cite{Zheng} jointly considered VM consolidation and traffic consolidation together to minimize the servers and network energy consumption in data centers. The authors not only model the server power model, but also the switch model in the network. Experiments conducted under real environment show that this joint approach outperforms the approaches that only adopt VM consolidation in energy consumption and service delay. Ferdaus et al. \cite{Ferdaus} proposed a VM consolidation algorithm combining with Ant Colony Optimization, in which a number of artificial ants select feasible solutions and exchange information for their solutions quality to obtain an optimized solution. As the authors consider multiple resource types, the VM consolidation process in this work is modeled as a multi-dimensional vector packing process.

The difference of DVFS and VM consolidation lies in that DVFS achieves energy saving through adjusting frequencies of processors rather than using less active servers. The DVFS approach introduces a trade-off between energy consumption and computing performance, where processor lowers the frequency/voltage when it is lightly loaded and utilizes full frequency/voltage when heavily loaded. 

Kim et al. \cite{Kim} modeled real-time service as real-time VM requests. To balance the energy consumption and price, they proposed several DVFS algorithms to reduce energy consumption.  Pietri et al. \cite{Pietri} introduced another energy-aware workflow scheduling approach using DVFS and its objective is finding an available frequency to minimize energy consumption while ensuring user deadline. Deng et al. \cite{Deng} coordinated CPU and memory together to investigate performance constraints, which is the first trial to consider them together when applying DVFS. They aim to find the most energy efficient frequency while ensuring system performance constraints. 

To reduce energy consumption, an approach \color{black} that \color{black} combines DVFS and VM consolidation together \color{black} was \color{black} presented in \cite{Teng}. The authors proposed several heuristic algorithms for batch-oriented scenarios. A DVFS-based algorithm for consolidating VMs on hosts is introduced to minimize the data center energy consumption while ensuring \color{black} Service Level Agreement of jobs. \color{black} The results demonstrate that these two techniques can work together to achieve better energy efficiency.

VM consolidation and DVFS have been proven to be efficient to reduce energy consumption, however, both of them cannot function well when the whole data center is overloaded. Therefore, we introduce a paradigm, called brownout, to handle data center overloads and reduce energy consumption. Originally, the brownout is applied to prevent blackouts through voltage drops in case of emergency. In Cloud scenario, it is first borrowed in \cite{Klein} to design more robust applications under the overloaded or unpredicted situation. Tomas et al. \cite{tomas} introduced a combined brownout-overbooking approach to improve resource utilization while ensuring response time. In our previous work, we \color{black} applied \color{black} brownout to save energy consumption in data centers.  In \cite{Xu}, we presented the brownout enabled system model and proposed several heuristic policies to find the microservices or application components that should be deactivated for energy saving purpose. We also introduced that there was a trade-off between energy consumption and discount in our model. In \cite{xu2017energy}, we extended our previous work and adopted approximate Markov Decision Process to improve the aforementioned trade-off.  Both in \cite{Xu} and \cite{xu2017energy}, the experiments are conducted under simulation environments. \color{black} Different from them\color{black}, in this paper, we implement a prototype system based on real infrastructure.  

\color{black}
Some other works related to energy-aware resource scheduling in Clouds are also proposed in the literature. Gai et al. \cite{GaiTCC} presented a cost-aware heterogeneous cloud memory model to provision memory services and considered energy performance. In \cite{GaiJPDC}, the authors introduced a novel approach that aimed to reduce the total energy cost of heterogeneous embedded systems in mobile Clouds. A dynamic energy-aware model to reduce the additional power consumption of wireless communications in the dynamic network environment was introduce in \cite{Gai}. Different from our work, these articles are not focused on data center energy consumption. 

\color{black}
In this work, our objective is reducing data center energy consumption while ensuring Quality of Service (QoS).  \color{black} Some related work considering power and QoS have also been conducted\color{black}. Khanouche et al. \cite{Khanouche} proposed an energy-aware and QoS-aware service selection algorithm, which is designed to solve a multi-objective optimization problem. But it is applied to the Internet of Things rather than data centers. Wang et al. \cite{Wang} used an improved particle swarm optimization algorithm to develop an optimal VM placement approach involving a tradeoff between energy consumption and global QoS guarantee for data-intensive services in national cloud data centers.  

Different from the energy efficient approaches based on VMs, our implementation is based on containers. Compared with VMs, containerization provides cloud application management based on lightweight virtualization. Currently, most work related to containers are focused on the orchestration of containers construction and deployment \cite{Pahl}. A detailed comparison of related work is shown in Table 1. 

To the best of our knowledge, our work is the first prototype system to reduce energy consumption with brownout based on containers, which also considers the trade-offs between energy consumption and QoS. Our prototype system provides practice and experience for finding complementary option apart from VM consolidation and DVFS.


\section{Motivations: scenarios and challenges}
\color{black} To study service providers' requirement and concerns for managing services based on containers, we give a motivation example of a real-world case study with brownout technology. 

A good example of the container-based system is the web-based service. An online shopping system implemented with containers are presented in \cite{Weaveshop}, which contains multiple microservices, including user, user database, payment, shipping, front-end, orders, carts, catalog, carts database and etc. As it is implemented with microservices, each microservice can be activated or deactivated independently. 
When requests are bursting, the online shopping system may be overloaded, and it cannot satisfy QoS requirements. To handle the overloads and reduce energy consumption, the brownout approach can be applied to temporarily disable some microservices, such as the recommendation engine, to save resource and power. 
By deactivating the recommendation engine, the system is able to serve more requests with the essential requirement and satisfy QoS. When the system is not overloaded anymore, the disabled microservices are activated again. 
Considering the overloaded situation, we assume that the service provider of this online shopping system is interested to improve QoS and save energy costs. In addition, the service provider may prefer to apply brownout to manage microservices in their systems. For such deployment, the service provider faces several challenges as below: \color{black}

\textbf{1. How to predict the tendency of future workload. }
It is common for cloud data centers meeting unexpected loads, which may lead overloaded situation and performance degradation. Estimating the workloads precisely enables the service providers to select proper resource management policy. 

\textbf{2. When to disable microservices. } 
Microservices can be dynamically deactivated or activated according to system conditions. A crucial decision should be made in both situations to determine the best time to deactivate containers to relieve overloads and reduce energy consumption while ensuring predefined QoS constraints. 

\textbf{3. \color{black} Which microservice to disable. \color{black} }  
Firstly, mandatory and optional microservices are required to be identified. The mandatory microservices, like the database, \color{black} must be kept \color{black} running all the time. While the optional microservices are allowed to be deactivated temporarily, such as the recommendation engine in the online shopping system. Secondly, once brownout is triggered, it may require selecting one or more microservices to deactivate. The challenge lies in determining the proper combinations of deactivated microservices to achieve the best beneficial results. 

\textbf{4. When to turn the hosts on or into low-power mode.} 
To reduce energy consumption, it is required to combine brownout and dynamically turning hosts into low power states, which saves the energy of idle hosts. To ensure QoS, it is also essential to determine efficiently when the host states should be switched, because hosts are required to be turned on quickly when requests are increasing. 

\textbf{5. How to design scheduling policy based on brownout.} 
In brownout-compliant microservices, there is a control knob called dimmer that represents a certain probability and shows how often the optional components are executed. It is required to design the dimmer value to be efficiently computed, which supports the brownout to be triggered quickly. The designed policy is also needed to be available for different preferences, like investigating the trade-offs between energy consumption and QoS.

To address aforementioned issues and enable system deployment based on containers and brownout, \color{black} we introduce our approach: \color{black} iBrownout. 

\section{iBrownout Architecture }
The architecture of iBrownout is demonstrated in Fig. 1 and its main components are explained below: 

 \begin{figure}[!ht]
	\centering
	\includegraphics[width=1.0\linewidth]{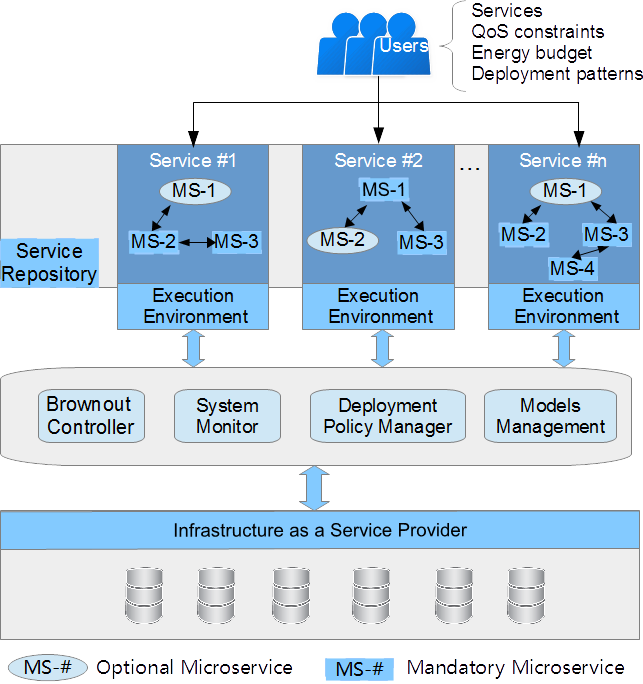}
	\caption[VarPerOptCom]{iBrownout Architecture}
	
	\label{fig:VarPerOptCom}
\end{figure}

1)    \textbf{Users:} All services provided by the system are available for users to submit their requests to cloud data centers. The user component contains user' information and requested services. In addition, the system administrator is also included in this component, in which it captures administrators' configurations such as predefined QoS constraints (including maximum response time, error rates and etc.), energy budget and service deployment patterns (in Docker, it is represented as a compose file \cite{Docker}). 

2)     \textbf{Cloud Service Repository:} The services provided by the service provider are managed by Cloud Service Repository component, which contains the service information, including service's name and image.  Each service may be constructed by several microservices, for example, in the online shopping system, the carts service manages items in user's cart, which contains cart microservice showing items in carts and cart database microservice storing items information. To manage microservices with brownout, the microservices are identified as mandatory or optional. 

\textbf{a.    Mandatory microservices:} The mandatory microservice keeps running all the time when it is launched, such as database-related microservices. 

\textbf{b.    Optional microservices:} The optional microservices are allowed to be activated or deactivated according to system status. Optional microservices have parameters like CPU utilization $u(MS_c)$, which indicates the amount of CPU usage when it is running and the reduced amount of CPU usage if it is deactivated. 

\textbf{3)    Execution Environment:} It represents the running environment for containerized applications. The dominant environments are Docker, Kubernetes and Mesos. In our prototype system, we adopt Docker to provide the execution environment for containers/microservices.  

\textbf{4)    Brownout Controller:} The operation of optional microservices are controlled by Brownout Controller, which determines operations based on system overloaded status. The Brownout Controller takes advantage scheduling policies that are introduced in Section 6 (Scheduling Policies) to offer an elegant solution for operating optional microservices.  It is also responsible for monitoring the health of all services. To adapt to our architecture, our dimmer in Brownout controller is different from the one in \cite{Klein} that requires a dimmer per application. Our dimmer is only applied to the optional microservices. Moreover, rather than based on response time, our dimmer is computed according to the severity of overloaded hosts (the number of overloaded hosts).

\textbf{5)    System Monitor:} These components provide health monitoring of nodes and collects hosts resource usage information. Third party monitoring toolkit can be used to provide a view of host status. For instance, the APIs provided by Grid'5000 \color{black} \cite{Grid5000} (a real cluster infrastructure in France)  \color{black} give users real-time reports on infrastructure metric, including host healthy \color{black} status\color{black}, utilization and energy consumption.

\textbf{6)    Scheduling Policy Manager:} This component provides a set of scheduling policies for Brownout Controller to schedule containers/microservices. Because there exist energy consumption budget and QoS constraints, we have to design and implement policies targeting for different preferences. For example, when service provider cares more about QoS, a scheduling policy that focuses on optimizing QoS will be applied.

\textbf{7)    Models Management:} It provides energy consumption and QoS models for the system. The power consumption model should be modeled to be relevant to microservice/container utilization, and the QoS model identifies the constraints of QoS. such as response time and error rates.

\textbf{8)    Cloud Infrastructure:} In infrastructure as a service model, Cloud providers offer bare metal to support service requests, which host multiple containers/microservices.  We take advantage of Grid'5000 clusters as our infrastructure.


In order to realize the proposed architecture, several techniques are utilized. 

Java: iBrownout is built using Java and it benefits from Java's feature to run on any platform with Java Virtual Machine. Components including Brownout Controller, System Monitor, Deployment Policy Manager and Models Management are all implemented with Java. These components \color{black} calls \color{black} Docker APIs to collect containers information, such as utilization of containers. 

Docker \cite{DockerDoc}: iBrownout takes advantage of Docker Swarm cluster to manage the containers/microservices, including microservices deployment, stop, start, update and etc.  Docker compose file is used to define features of containers, such as whether containers are optional, which containers are deployed, how many containers are provided, how much resources are allocated to containers, deployment constraints of containers and dependencies between different containers.  

Ansible \cite{Ansible}: It is a toolkit to automate applications provisioning, configuration management and application deployment. iBrownout utilizes it to send management operations among nodes.

\section{Modelling and Problem Statement}

In this section, we will introduce the models in our system and state the problem we aim to optimize. Table 2 presents the symbols and their \color{black} meanings \color{black} used in this paper. \color{black} For example, we use $h_i$ to denote host $i$ and $P_i(t)$ to represent the power of $h_i$ at time interval $t$. \color{black}

 \begin{table}[]
	
	\centering
	\color{black}
	\caption{Symbols and their meanings}
	\color{black}
	\label{my-label}
	\resizebox{0.48\textwidth}{!}{%
		\begin{tabular}{|c|l|}
			\hline
			\textbf{Symbols}           & \multicolumn{1}{c|}{\textbf{Meanings}}                 \\ \hline
			$h_i$                     & Server (host) $i$                                        \\ \hline
			$t$                       & Time interval $t$                                        \\ \hline
			$P_i(t)$            & Power of  $h_i$ at time $t$                             \\ \hline
			$P_i^{idle}$              & Power when  $h_i$ is idle                                \\ \hline
			$P_i^{dynamic}$           & Power when  $h_i$ is fully loaded                        \\ \hline
			$P_i^{max}$               & Maximum power of  $h_i$                                  \\ \hline
			$hl$                      & Server list in data  center                               \\ \hline
			$M$                       & Size of server list $hl$ \\ \hline
			$N_i$                     & Number of microservices assigned to  $h_i$                         \\ \hline
			
			$u_i$                     & Utilization of host $h_i$
			\\ \hline
			
			$MS_{i, j}$               & Microservice $j$ on  $h_i$                                         \\ \hline
			$u(MS_{i, j})$            & Utilization of microservice $j$ on  $h_i$                          \\ \hline
			$E(t)$                    & Energy consumption at time interval $t$ \\ \hline
			
			$u_t$                     & Overloaded threshold of host \\ \hline
			
			$OTR(u_t)$                & Overloaded time ratio according to $u_t$ 
			\\ \hline
			
			$k$                      & Maximum percentile value of response time 
			\\ \hline
			
			$t_v$                    & Time threshold of SLA violation
			\\ \hline
			
			$SLAVR(t_v)$                 & SLA violation ratio according to violation time threshold $t_v$ \\ \hline
			
			$Num_v$                   & The number of requests that violate SLA
			\\ \hline
			
			$Num_a$                   & The total number of requests from clients
			\\ \hline

			$C$                       & The maximum number of containers on hosts
			\\ \hline
			
			$\alpha$                  & The maximum allowed overloaded time ratio
			\\ \hline
			
			$\beta$                   & The maximum allowed average response time
			\\ \hline
			
			$\phi$                    & The maximum allowed 95th percentile of response time
			\\ \hline
			
			$\gamma$                  & The maximum allowed SLA violation ratio
			\\ \hline
			
			$M_a$                     & The number of current active hosts
			\\ \hline
			
			$M_a^{'}$                   & The updated number of active hosts for Auto-scaling policy
			\\ \hline
			
			$n_o$                    & Overloaded threshold of request number based on profiling data
			\\ \hline
			
			$n_r$                   & Request rate
			\\ \hline
			
		    $ocl_{i,t}$             & The optional container/microservice list on $h_i$ at time interval $t$
		    \\ \hline
		    
		    $\mathbb{P}(ocl_{i, t})$  & The power set of $ocl_{i, t}$  
		    \\ \hline
		    
		    $dcl_{i,t}$               & The deactivated container/microservice list on $h_i$ at time interval $t$
		    \\ \hline
		    
		    $HUM()$                   & Host utilization model to compute host power based on host utilization
		    \\ \hline
		    \color{black}
		    $HP$ & \color{black} The expected host power calculated by host utilization model
		    \\		    \hline
		    
		    $TP$                   &  The overloaded power threshold 
		    \\ \hline
		    
		    $u_i^r$               & The expected utilization reduction
		    \\ \hline
		    
		    $u(dcl_{i,t})$        & The utilization of deactivated container/microservice list   
		    \\ \hline
		    
		    $n_t$                 & The number of overloaded hosts at time interval $t$
		    \\ \hline
		    
		    $\theta_t$            & The dimmer value
		    \\ \hline
		    
		    $COH()$                 & Compute overloaded hosts
		    \\ \hline
		    
		    $HPM()$                & Host power model to compute host utilization based on host power
		    \\ \hline
		    
		    $P_i^r$                & Expected power reduction of $h_i$
		    \\ \hline
		    
		    $MS_c$                 & Container/microservice $c$
		    \\ \hline
		    
		    $S_t$                  & The set of deactivated containers/microservice connection tags
		    \\ \hline
		    
		    $Ct(MS_c)$             & Connection tag of $MS_c$
		    \\ \hline
		    
		    $X$                    & Random variable to generate sublist of $ocl_{i,t}$
		    \\ \hline
		    
		\end{tabular}%
	}
\end{table}

\subsection{Power Consumption}
We adopt the servers power model derived from \cite{Zheng}. The power of server $i$ is $P_i(t)$ that is dominated by the CPU utilization:
\begin{equation}
P_i(t) = 	\begin{cases}
P_i^{idle} +  u_i \times P_i^{dynamic}   & ,N_i > 0\\
0 & ,N_i = 0
\end{cases}
\end{equation}

$P_i(t)$ is composed of idle power and dynamic power. The idle power is regarded as constant and the dynamic power is linear to the server utilization $u_i$ \cite{Zheng}. If no container or microservice is hosted on a server,  the server is turned off to save power. The server CPU utilization equals to  total CPU utilization of all the containers/microservices deployed to the server, which is represented as:
\color{black}
\begin{equation}
u_i = \sum_{j=1}^{N_i}u(MS_{i, j}(t))
\end{equation}
\color{black}
where $MS_{i,j}$ refers to the $j$th microservice on server $i$, $N_i$ represents the number of microservices deployed to server $i$. And $u(MS_{i,j}(t))$ refers to the CPU utilization of the container/microservice at time interval $t$.

Then the total energy consumption during time interval $t$, with $M$ servers is:

\begin{equation}
E(t) = \sum_{i=1}^{M}\int_{t-1}^{t} P_{i}(t) dt
\end{equation}

\subsection{Quality of Service}

To model the QoS requirement in our system, we adopt several QoS metrics as below:

\textbf{Overloaded Time Ratio:} based on host loads, we define two states for hosts: overloaded and non-overloaded. Overloads will lead hosts to experience performance degradation. We regard host as overloaded when host utilization is above the predefined utilization threshold. To evaluate this QoS metric to be independent of workloads, we adopt the metric introduced in \cite{AntonTPDS}, which is denoted as Overloaded Time Ratio (OTR):
\begin{equation}
OTR(u_t) = \frac{t_o(u_t)}{t_a}
\end{equation}
where $u_t$ is the overloaded CPU utilization threshold; $t_o$ is the time period that host is identified as overloaded, which is relevant to $u_t$; and $t_a$ is the total time periods of the hosts. As a QoS constraint, this metric is configured as the maximum allowed value of $OTR$. For instance, if the system SLA is defined as 10\%, it the time period of overloaded states for all the hosts is less then 10\%. The SLA constraint can be formulated as: 
\color{black}
\begin{equation}
\frac{1}{M}\sum_{i=1}^{n=M}OTR_{n}(u_t) \leq 0.1
\end{equation}
\color{black}
where $M$ is the total number of hosts in the data center. As introduced the later, our brownout-based approach checks the host status at each time period and triggers the brownout to deactivate when there are overloaded hosts. Therefore, this metric also represents the ratio that brownout is triggered. 

\textbf{Response time:} this metric measures the time that from sending requests to receiving requests. We also evaluate the response time with the maximum of $k$th percentile response time of all requests, where $k$ could be 90, 95, 99 and etc. For example, if the maximum of 95th percentile response time equals to 1 second, it means that 95\% of all requests get the response within 1 second.

\textbf{SLA Violation Ratio:} It represents how many requests are failed due to overload. If clients send $Num_a$ requests to the system, and $Num_{err}$ of them are returned with errors, then error rate is represented as:
\color{black}
\begin{equation}
SLAVR = \frac{Num_{err}}{Num_{a}}
\end{equation}
\color{black}

\subsection{Optimization Objective}
As discussed in the previous section, it is necessary to minimize the total energy consumption, while ensuring QoS by avoiding overloads, decreasing response time and reducing error rates. 
Therefore, our problem can be formulated as an optimization problem (7)-(10):
\begin{equation}
\min{\sum_{t=1}^{T}E(t)}
\end{equation}
\color{black}
\begin{equation}
\frac{1}{M}\sum_{n=1}^{n=M}OTR_{n}(u_t) \leq \alpha
\end{equation}
\color{black}
\begin{equation}
R_{avg}^t \leq \beta, \ R_{95th}^t \leq \phi
\end{equation}

\begin{equation}
SLAVR \leq \gamma
\end{equation}
where $\sum_{t=1}^{T}E(t)$ is the total energy consumption of data center, $\alpha$ is the maximum allowed average response time of overloaded states; $R_{avg}^t$ is the average response time and $\beta$ is the allowed average response time;  $R_{95th}^t$ is the maximum of 95th percentile response time and $\phi$ is the allowed the 95th percentile response time, and $\gamma$ is allowed SLA violation ratio. 

\section{Scheduling Policy}
In this section, we will introduce our brownout-based \color{black} scheduling \color{black} policies. Prior to brownout approach, we require an auto-scaling algorithm to dynamically add or remove hosts to utilize host resource more efficiently. 
\subsection{Auto-scaling Policy}

\begin{algorithm}
	\color{black}
	\footnotesize
	\caption{Auto-scaling Policy}
	\begin{algorithmic}[1]
		
		\renewcommand{\algorithmicrequire}{\textbf{Input:}}
		\renewcommand{\algorithmicensure}{\textbf{Output:} }
		\REQUIRE  host list $hl$ with size $M$, number of active hosts $M_a$, number of requests when  host is overloaded $n_o$, recent request rate in the recent time $n_r$. 
		\ENSURE   number of active hosts $M_{a'}$ 
		\\ 
		\STATE $M_a$ $\leftarrow$ number of current active hosts 
		\STATE $n_o$ $\leftarrow$ overloaded threshold of request number according to profiling data 
		\STATE $n_r$ $\leftarrow$ number of request rate at current time window according to previous time windows
		\STATE $M_{a'} \leftarrow \lceil n_r \div n_o \rceil $    
		\STATE $M'  \leftarrow M_{a'} - M_{a} $

		\IF{$M' > 0$}
		\STATE  Add $M' hosts$
		\ELSIF{$M' < 0$}
		\STATE Remove $|M'| hosts$
		\ELSE
		\STATE no scaling
		\ENDIF

		\STATE update number of active hosts with $M_{a'}$
		
	\end{algorithmic} 
\end{algorithm}

 We adopt the auto-scaling algorithm in \cite{Toosi}, which is a predefined threshold-based approach. With profiling experiments, we configure the requests overloaded threshold above which the host cannot respond \color{black} to \color{black} requests within an acceptable time limit. As shown in Algorithm 1, in the initialization stage, the master node that runs auto-scaling algorithm firstly gets the number of current active hosts (line 1), sets the overloaded threshold of request number according to profiling data (line 2) and fetches the request rate at current time window according to previous time windows (line 3). The advantage of sliding time window is to give more weights to the values of recent time windows, and more details will be given in Section 7. Line 4 shows the method to compute the current required hosts $M_{a'}$, which is the ratio of current request rate and the overload threshold. If the required number of hosts is more than current active hosts, more hosts will be added to provide services, otherwise, if current active hosts are more than required, then the excess machine can be set as low-power mode to save energy consumption (lines 6-12). Finally, the master node will update the number of active hosts. 

\subsection{Initial Deployment}

\begin{figure}[!ht]
	\centering
	\includegraphics[width=1.0\linewidth]{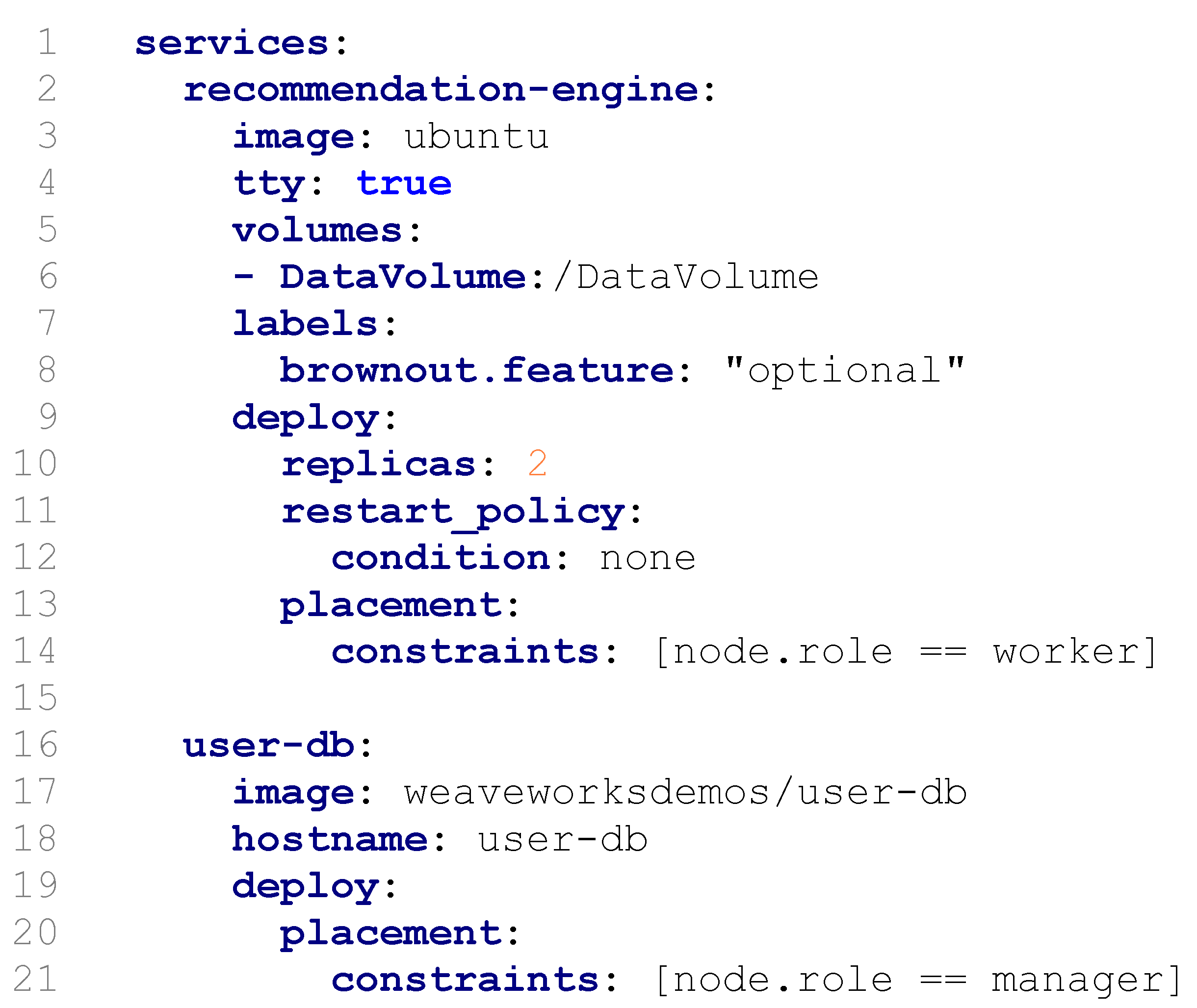}
	\caption[VarPerOptCom]{Simple example of Docker compose file}
	
	\label{fig:VarPerOptCom}
\end{figure}
In the initial deployment stage, containers are deployed based on Docker compose file, which identifies the all the required information of services and the configurations of initial deployment. A simple example is shown in Fig. 2. Lines 2-14 show the information of recommendation engine service, which is built on the Ubuntu image and attached with a data volume. The recommendation engine is set as optional microservice, which can be deactivated and has two replicates. Moreover, this service will only be deployed on Docker worker node as deployment constraint. Lines 16-21 demonstrate the information of user database service, which is not optional and restricts to be deployed to Docker master node. 

\subsection{Optimization Deployment with Scheduling Policies based on Brownout}

We have proposed three brownout-based policies as follows:

\subsubsection{\textbf{Lowest Utilization Container First (LUCF)}} 

 \begin{algorithm}
	\color{black}
	\footnotesize
	\caption{Lowest Utilization Container First Policy (LUCF)}
	\begin{algorithmic}[1]
		
		\renewcommand{\algorithmicrequire}{\textbf{Input:}}
		\renewcommand{\algorithmicensure}{\textbf{Output:} }
		\REQUIRE  host list $hl$ with size $M$, microservice information, overloaded power threshold $TP$, dimmer value $\theta_t$ at time $t$, scheduling interval $T$, deactivated component list $dcl_{i, t}$ on host $h_i$, power model of host $HPM$,   the optional component list $ocl_{i, t}$, which is sorted based on utilization $u(MS_c)$ in ascending order
		\ENSURE   total energy consumption, number of shutting down hosts 
		\\ 
		\STATE initialize parameters with inputs, like $TP$  

		\FOR {$t \leftarrow 0$ to $T$}
		
		\STATE $n_t \leftarrow COH(hl)$
		\IF{$n_t > 0$}
		\STATE  $\theta_t$ $\leftarrow$ = $ \sqrt{\frac{n_t}{M}}$
		
		\FORALL {$h_i$ in $hl$  (i.e. $i = 1, 2, \dots, M$) }
		
		\IF{($P_{i}(t)$ $\textgreater$ $ TP$)}
		\STATE $P^{r}_i$ $\leftarrow$ $\theta_t$ $\times$ $P_{i}(t)$
		\STATE $u_{i}^{r}$ $\leftarrow$ $HPM$($h_i, P^{r}_i$)
		\STATE $dcl_{i, t}$ $\leftarrow$ NULL
		\STATE  $S_{t}$$\leftarrow$ NULL
		
		\IF{$u(MS_{1})$ $\geq$ $u_{i}^{r}$}
		\STATE  $dcl_{i, t}$ $\leftarrow$  $dcl_{i, t}$ + $MS_{1}$
		\STATE  $S_{t}$$\leftarrow$ $S_{t}$ + $Ct(MS_{1})$
		\ENDIF
		
		\FOR {$MS_c$ in $ocl_{i, t}$}
		\IF{$(u(MS_{c})$ $\leq$ $u_{i}^{r})$ \& $(u(dcl_{i, t}) \leq u_{i}^{r})$}
		\STATE  $dcl_{i, t}$ $\leftarrow$  $dcl_{i, t}$ + $MS_{c}$
		\STATE  $S_{t}$$\leftarrow$ $S_{t}$ + $Ct(MS_{c})$
		\color{black}
		\STATE  $ \min \leftarrow   (u_{i}^{r} - u(dcl_{i, t}))$ 
		\color{black}
		\ENDIF
		\ENDFOR
		
		\FORALL {$MS_c$ in $ocl_{i, t}$}

		\IF{$Ct(MS_c)$ in $S_t$}
		\STATE $dcl_{i, t} \leftarrow dcl_{i, t} + MS_c$
		\ENDIF
		\ENDFOR
		
		\ENDIF
		\STATE deactivate components in $dcl_{i, t}$
		\ENDFOR
		\ELSE 
		\STATE activate deactivated components
		\ENDIF

		\ENDFOR

	\end{algorithmic} 
\end{algorithm}

The Lowest Utilization Container First policy selects a set of containers with the lowest utilization that reduces the utilization to be less than the overloaded threshold of a host is overloaded. Let $ocl_{i, t}$ be the optional container list on host $h_i$ at time interval $t$. Let $\mathbb{P}(ocl_{i, t})$ to be the power set of $ocl_{i, t}$, the LUCF finds the deactivated container list $dcl_{i, t}$, which is included in  $\mathbb{P}(ocl_{i, t})$. \color{black} The deactivated container list minimizes the value difference between the expected utilization reduction $u_i^r$ and its utilization $u(dcl_{i, t})$ \color{black} The  deactivated container list is defined in Equation (11). 

	\color{black}
\begin{equation}
{\small 
dcl_{i, t} = \begin{cases}
\{ HP \leq TP, 
u_{i}^r -  u(dcl_{i, t}) \rightarrow \min \},  & if \ P_i(t) \geq TP \\
\emptyset, & if  \ P_i(t) < TP
\end{cases}
}
\end{equation}
\color{black}
\color{black} where $HP$ is the expected host power calculated by host utilization model $HUM(h_i, u_{i} - u(dcl_{i, t}))$ that fetches the host power based on host utilization $u_i - u(dcl_{i, t})$; \color{black} $TP$ is the overloaded power threshold of $h_i$. 

The pseudocode of LUCF is shown in Algorithm 2, which mainly consists of 8 steps as discussed below. Before entering the approach procedures, service provider firstly needs to initialize input parameters for the algorithm, such as overloaded power threshold (lines 1-2). The power threshold $TP$ is a value for checking whether a host is overloaded. 

1) In each time interval $t$, checking all the hosts status and counting the number of overloaded hosts as $n_t$ (line 3).

2) Adjusting the dimmer value $\theta_t$ as $\sqrt{ \frac{n_t} {M}}$ based on the number of overloaded hosts $n_t$ and host size $M$ (line 5). As introduced in related work, the dimmer value $\theta_t$ is applied to compute the adjustment degree of power consumption at time $t$. The dimmer value $\theta_t$ is 1.0 if all the hosts are overloaded at time $t$ and it means that brownout controls containers/microservice on all the hosts. The dimmer value is 0.0 if no host is overloaded and brownout will not be triggered at time $t$. The \color{black} adjustment of dimmer \color{black} presents that the dimmer value is relevant to the number of overloaded hosts.

\begin{table}[]
	\centering
	\caption{Power consumption of selected node at different utilization levels in Watts}
	\label{my-label}
	\begin{tabular}{|c|c|c|c|c|c|c|}
		\hline
		\textbf{Utilization}   & \multicolumn{1}{l|}{Sleep} & \textbf{0\%}  & \textbf{10\%} & \textbf{20\%} & \textbf{30\%} & \multicolumn{1}{l|}{\textbf{40\%}} \\ \hline
		Power (Watts) & 10                         & 201  & 206  & 211  & 213  & 216                       \\ \hline
		\textbf{Utilization}   & \textbf{50\%}                       & \textbf{60\%} & \textbf{70\%} & \textbf{80\%} & \textbf{90\%} & \textbf{100\%}                     \\ \hline
		Power (Watts)   & 221                        & 223  & 225  & 231  & 233  & 237                       \\ \hline
	\end{tabular}
\end{table}

3) Calculating the expected utilization reduction on the overloaded hosts (lines 7-9). Based on the dimmer value and host power model, LUCF calculates expected host power reduction $P_i^r$ (line 8) and expected utilization reduction $u_{i}^r$ (line 9) respectively. In our host power model, the host power consumption is mainly relevant to it CPU utilization. As shown in Table 3, we list power consumption at different CPU utilization levels of one host in Grid'5000 (Sagittare cluster in Lyon). In this power model, for example, the host with 100\% utilization is 237 Watts and 80\% utilization is 231 Watts, if the power is required to be reduced from 237 to 231 Watts, the expected utilization reduction is  $100\%-80\% = 20\%$.    

4) Resetting the deactivated container list $dcl_{i, t} $ and the set of deactivated container connection tags $S_t$ as empty (lines 10-11). This list and the set will be ready to collect deactivated containers and their connection tags.  

5) Finding the containers \color{black} to \color{black} be deactivated (lines 16-27). The LUCF  sorts the optional container list $ocl_{i, t}$ based on container utilization parameter in ascending \color{black} order \color{black}, therefore, the container with the lowest utilization is put in the head of the list. 
\color{black} Since we consider connected containers, each container has a connection tag $Ct(MS_c)$ that shows how it is connected with other containers.
If the first container utilization parameter value is above $u_{i}^r$, Algorithm 2 adds this container into the deactivated container list $dcl_{i, t}$ and  inserts its connection tag $Ct(MS_1)$ into $S_t$ (lines 12-13). 
\color{black} After that, Algorithm 2 finds other connected containers and adds them into deactivated container list (line 14).
If the first container utilization does not satisfy the expected utilization reduction, Algorithm 2 finds the containers sublist in the optional container list to deactivate more containers (lines 16-22). 
The utilization of this sublist is closest to the expected utilization reduction among all the sublists.  

Algorithm 2 also puts all the containers in the sublist into the deactivated containers list and puts their connection parameters into the $S_t$\color{black}. \color{black} 
\color{black} For connected containers, the sorting process is modified as treating the connected containers together for sorting, which lowers the priority of deactivating the connected containers, and avoids deactivating too many containers due to connections. \color{black}

6) Finding other connected container and puts them into the deactivated container list (lines 23-27).

7) Deactivating the containers in the deactivated container list (line 29).

8) In algorithm 2, if no host is above the power threshold, the algorithm activates the deactivated containers (line 32).

It is noticed that when the whole data center is overloaded, auto-scaling cannot add more hosts because of the limited resource. LUCF takes effects when Auto-scaling cannot function well, to be more specific, LUCF can be embedded into line 7 in Algorithm 1 to handle with overloads and reduce energy consumption. 

 \textbf{Algorithm Complexity:} the complexity of LUCF at each time interval is calculated as below: 
 the complexity of finding the deactivated containers is $O(C*M) $, where $C$ is the maximum number of containers on hosts and $M$ is the number of hosts.  
 The complexity of finding the connected components is also $O(C*M) $. Therefore, the complexity at each time interval of LUCF is the sum of these parts, which is $O(2*C*M) $. To be noted, line 3 relies on the network connection, if $C$ and $M$ are small, the network delay $O(T_d)$ can be a dominant part of algorithm execution time. Please see the results in Section 7.4.

\subsubsection{\textbf{Minimum Number of Components First Policy (MNCF)}}
The Minimum Number of Containers First (MNCF) policy selects the minimum number of containers while reducing the energy consumption in order to deactivate fewer services, as formalized in Equation (12). We do not provide the pseudocode of MNCF here because it is quite similar to the LUCF algorithm introduced earlier. 
\color{black}
\begin{equation}
{\small 
dcl_{i, t} = \begin{cases}
\{ HP \leq TP,   
|u(dcl_{i, t})| \rightarrow \min \},  & if \ P_i(t) \geq TP \\
\emptyset, & if  \ P_i(t) < TP
\end{cases}
}
\end{equation}
\color{black}

\subsubsection{\textbf{Random Selection Container Policy (RSC)}}
The Random Selection Container policy (RSC) policy takes advantage of a random selection of a number of optional containers to reduce energy consumption. Based on a uniformly distributed discrete random variable $(X)$, which selects randomly a subset of $dcl_{i, t}$, RSC is presented in Equation (13).

	\color{black}
\begin{equation}
{\small 
dcl_{i, t} = \begin{cases}
\{ HP \leq TP, 
X = U(0, |ocl_{i, t}| - 1) \},  & if \ P_i(t) \geq TP \\
\emptyset, & if  \ P_i(t) < TP
\end{cases}
}
\end{equation}
\color{black}

\section{Performance Evaluation}
We are evaluating our techniques experimentally on INRIA Grid'5000 testbed for Wikipedia web workload. \color{black} We also compare the performance with related policies introduced in \cite{Beloglazov}, \cite{tomas} and \cite{Toosi}. \color{black}

\subsection{Workload}

\begin{figure}[!ht]
	\centering
	\includegraphics[width=0.85\linewidth]{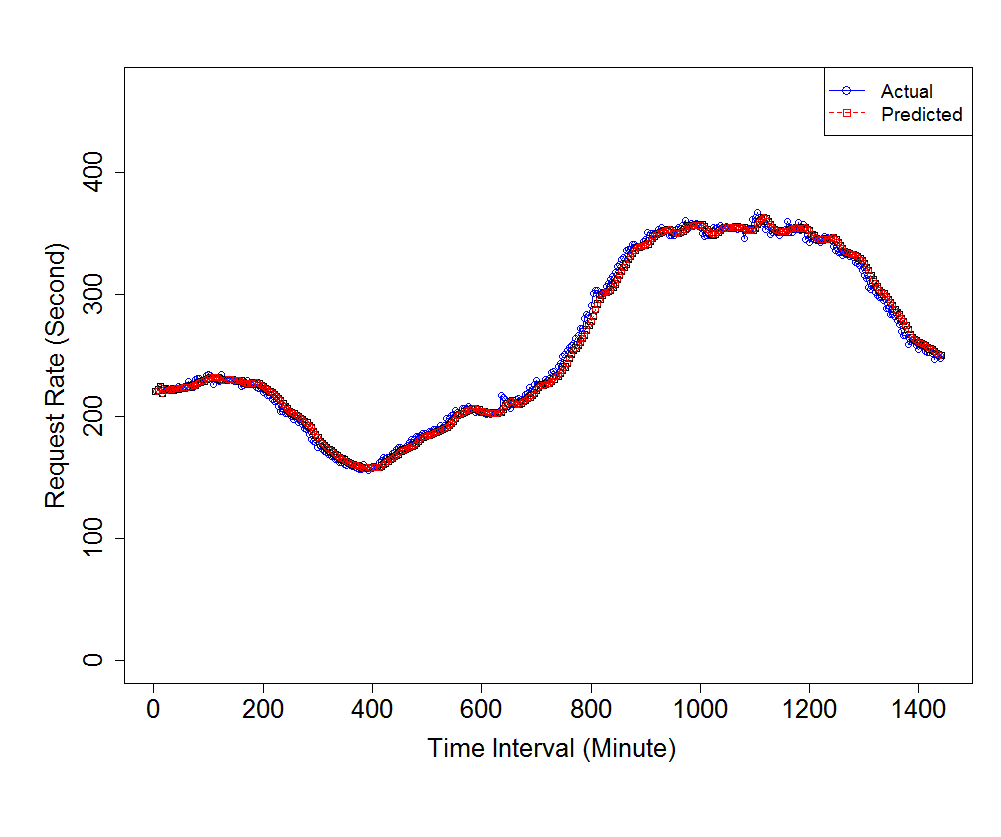}
	\caption[VarPerOptCom]{Predicted and actual requests rate}
	
	\label{fig:VarPerOptCom}
\end{figure}

We use real trace from Wikipedia requests on 2007 October 17 to replay the workload of Wikipedia users. To scale the workload set to fit with our experiments, we use 5\% of the original user requests size.  JMeter \cite{JMeter} is a toolkit designed for load testing and performance measurement, we use it to generate the requests by replaying the Wikipedia trace.  
 ${n_r}$ is the predicted request rate, which is calculated based on a sliding window \cite{AntonTPDS}.  Let $L_w$ to be the window size, and $n_r(t)$ to be the request rate at $t$, we estimate ${n_r}$ as:


\begin{equation}
{n_r}(L_w) = \frac{1}{L_w}\sum_{t=0}^{L_w-1} n_r(t)
\end{equation}
\color{black} In \color{black} our experiments, we set the sliding window size as 5. Fig. 3 shows the requests rate per second during the day, and the predicted rates and the actual rates are quite close.

\subsection{Testbed}

We use Grid'5000 \color{black} \cite{Grid5000}\color{black}, a French experimental grid platform, as our testbed. We adopt the cluster equipped with power measurement APIs at Lyon site, which is located at the southeast French. 
The architecture of prototype system deployed on the Grid'5000 clusters is presented in Fig.4, which shows that all the nodes are deployed with Docker swarm and categorized according to different roles as below:
\begin{itemize}

\item Master node: this node is initialized as the master node and running some services that can only be deployed on the master node, such as the brownout controller containing scheduling policies, as well as the Java Runtime and Ansible toolkit. 

\item Worker node: these nodes are workers that running services apart from the services on master node and database services. We have multiple worker nodes in our system. 

\item  Worker node (node only for the database): the database services are deployed on a specific worker node, which only hosts database-related services. 
\end{itemize}

We also have another node, namely request node, that contains workload trace and installed with JMeter to send requests to our cluster. This node can be located at any place to simulate users' behavior. In our experiments, to reduce the impacts of uncontrolled network traffic out of Lyon cluster, we also locate this node in Lyon cluster. 
 
 \color{black}
 The hardware information of our selected nodes is as below:
 \begin{itemize}

 \item Machine model: Sun Fire V20z. The maximum power of this model is 237 Watts, and its power of sleep mode is 10 Watts;
 \item Operating system: Debian Linux;
 \item CPU: AMD Operon 250 with 2 cores (2.4 GHz);
 \item Memory: 2 GB
\end{itemize}

 One of the nodes is running as the Docker Swarm master node, and other nodes are running as worker nodes. All required applications, such as Java, Docker, Ansible and JMeter, are installed in advance to minimize the impacts of CPU utilization and network traffics. 
 
  \color{black}
 
\begin{figure}[!ht]
	\centering
	\includegraphics[width=1.0\linewidth]{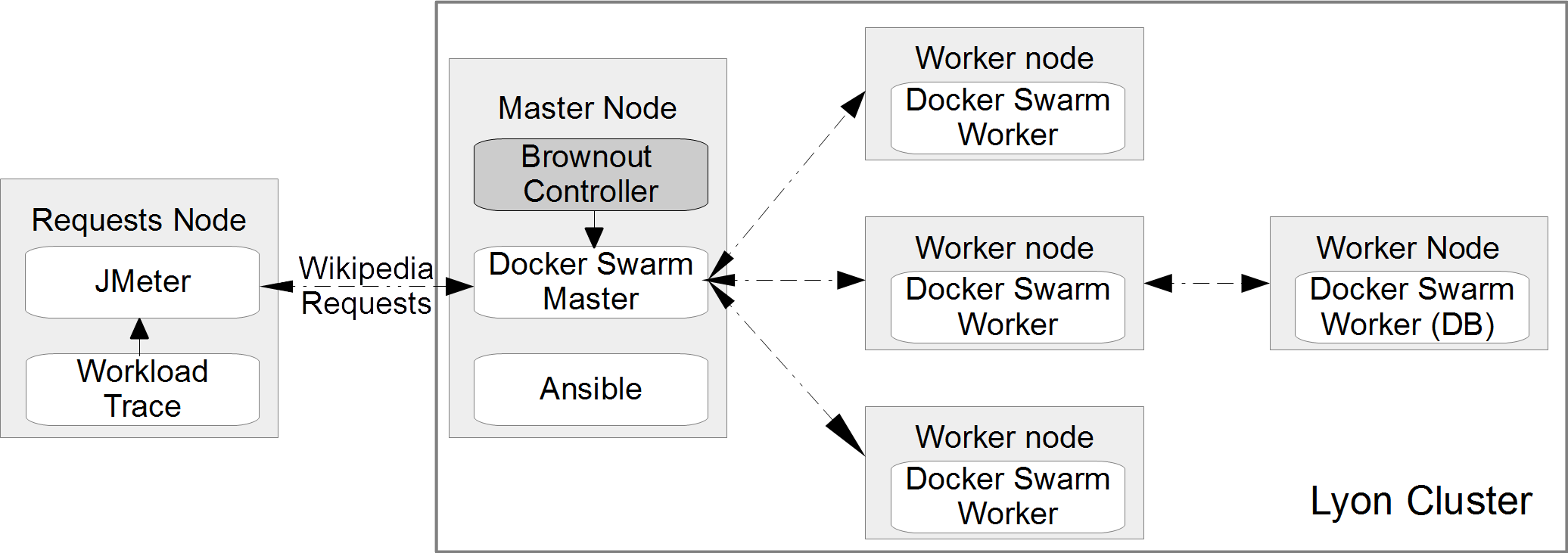}
	\caption[VarPerOptCom]{Architecture of Prototype System}
	
	\label{fig:VarPerOptCom}
\end{figure}

\subsection{Results}

\begin{figure*}[ht]
	\centering
	\includegraphics[width=0.9\linewidth]{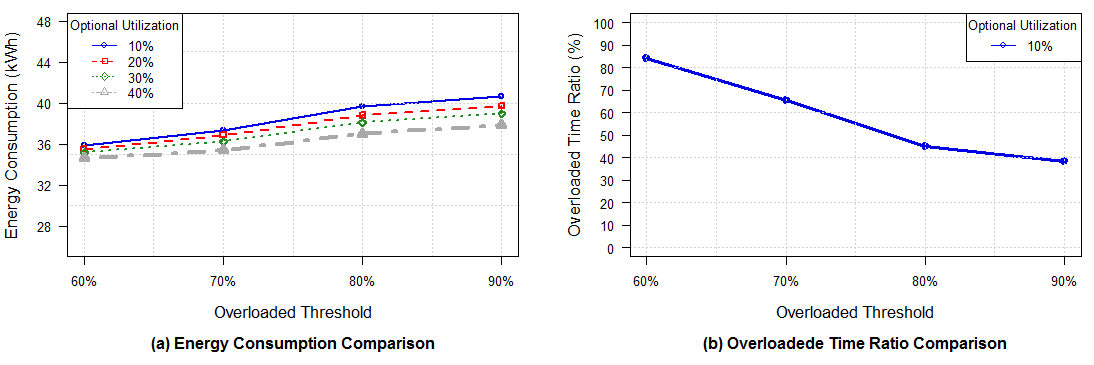}
	\caption[VarPerOptCom]{Algorithm Performance Comparison}
	
	\label{fig:VarPerOptCom}
	\end{figure*}

\color{black} To evaluate the performance of our proposed policies, we use three benchmark policies for comparison. 

1). \textbf{Non-Power-Aware (NPA)} policy \cite{Beloglazov}: it applies no power-aware optimization and hosts are keeping on all the time. We give 13 nodes as the resource for NPA. 

2). \textbf{Brownout-OverBooking (BOB)} policy \cite{tomas}: it aims to maximize actual utilization while reducing response time and minimally triggering brownout. The brownout operation in BOB is based on response time. When the response time is less than target utilization, the approach gradually increases application utilization. To let BOB  experience overloads, only 10 nodes are given to it. 

3). \textbf{Auto Scaling (Auto-S)} policy \cite{Toosi}: it dynamically scales in and out the number of active hosts as introduced in Algorithm 1. To let Auto-S endure overloads, we also give 10 nodes to Auto-S. 

For our proposed policies, they have the identical resource as BOB and Auto-S. In the following experiments, we mainly investigate two parameters: overloaded threshold and optional utilization percentage. \color{black}

\textbf{Overloaded threshold:} it represents the CPU utilization threshold that identifies whether a host is overloaded. We adopt this parameter \color{black} since \color{black} \cite{Beloglazov} have shown that it has an impact on energy consumption. It is varied from 60\% to 90\% in increments of 10\%. We choose this range because of the smaller overloaded threshold, like 50\%, means hosts are easier to be identified as overloaded and it will lead to inefficient resource usage.

\textbf{Optional utilization percentage:} it identifies how much CPU resource is given to optional containers, which also means how much CPU utilization can be reduced to save energy consumption. This parameter is investigated because \cite{Xu} shows that it influences the power consumption. \color{black} It \color{black} is varied from 10\% to 40\% in increments of 10\%. We choose these ranges because \cite{Xu} shows large optional utilization percentage, like 50\%, comes along much revenue loss and non-negligible experience degradation.
 
\textit{\textbf{(1) Comparison with different overloaded thresholds}}
\color{black}
\begin{figure*}[!ht]
	\centering
	\includegraphics[width=1.0\linewidth]{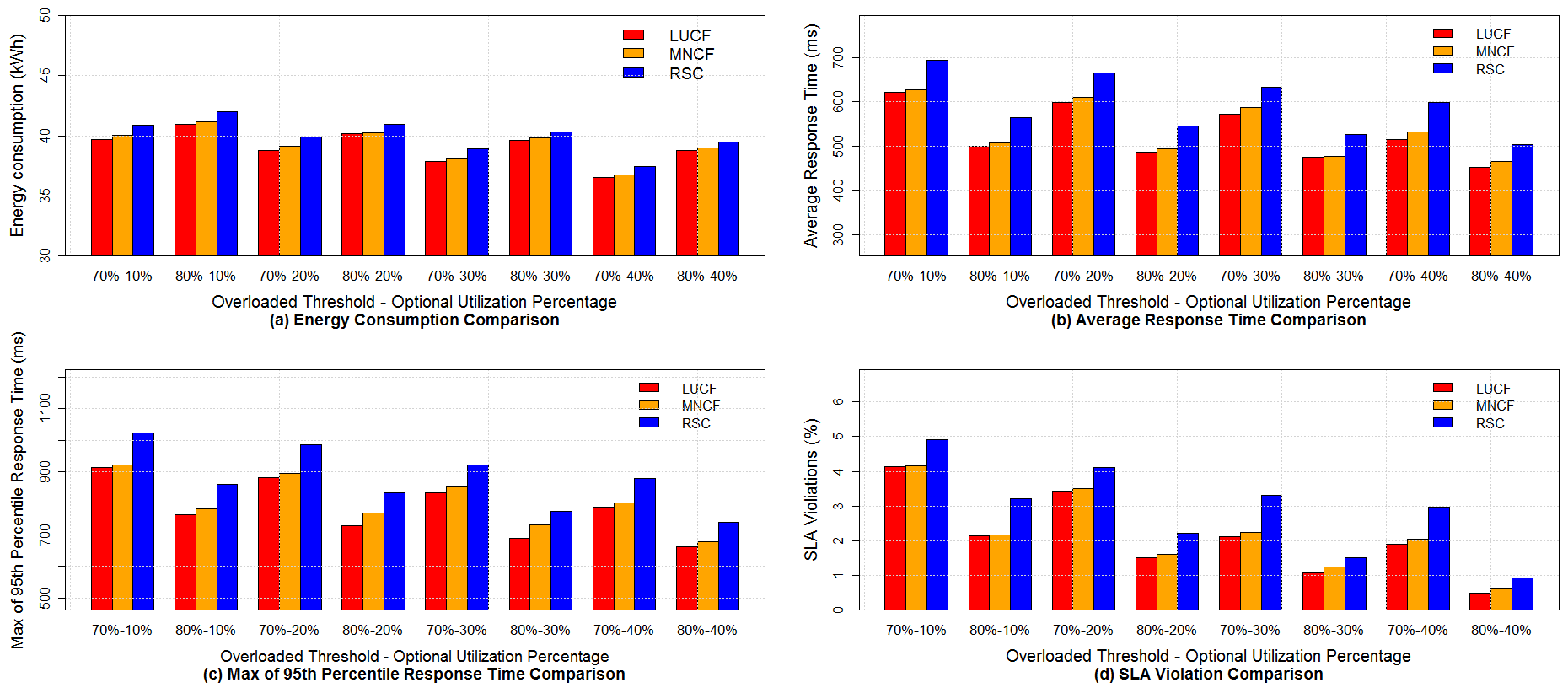}
	\caption[VarPerOptCom]{\color{black} Performance Comparison of Proposed Policies}
	
	\label{fig:VarPerOptCom}
\end{figure*}
\color{black}

 We have conducted several experiments with different values of overloaded threshold and optional utilization percentage for LUCF policy. \color{black} In Fig. 5, \color{black} the results show that when the overloaded threshold is higher, LUCF reduces less energy consumption, and when the system has higher optional utilization percentage, LUCF saves more energy consumption. However, as shown in Fig. 5 (b), when the overloaded threshold is smaller, like 60\%, the overloaded time ratio is quite high (around 85\%), which means hosts are regarded as overloaded in most time periods and brownout will be triggered frequently. As optional utilization percentage does not influence overloaded time ratio, we only show the LUCF with 10\% optional utilization here. \color{black} From the results, we observe a trade-off between energy consumption and overloaded ratio time when the overloaded threshold is varied, and we find out that configuring the overloaded threshold as 70\% and 80\% achieves better trade-offs, which reduces energy consumption while not triggering brownout too frequently. Therefore, we conduct experiments under 70\% and 80\% overloaded thresholds to compare our proposed policies in the following section. \color{black}

\textit{\textbf{(2) Comparison with proposed policies}}

\begin{figure*}[!ht]
	\centering
	\includegraphics[width=0.9\linewidth]{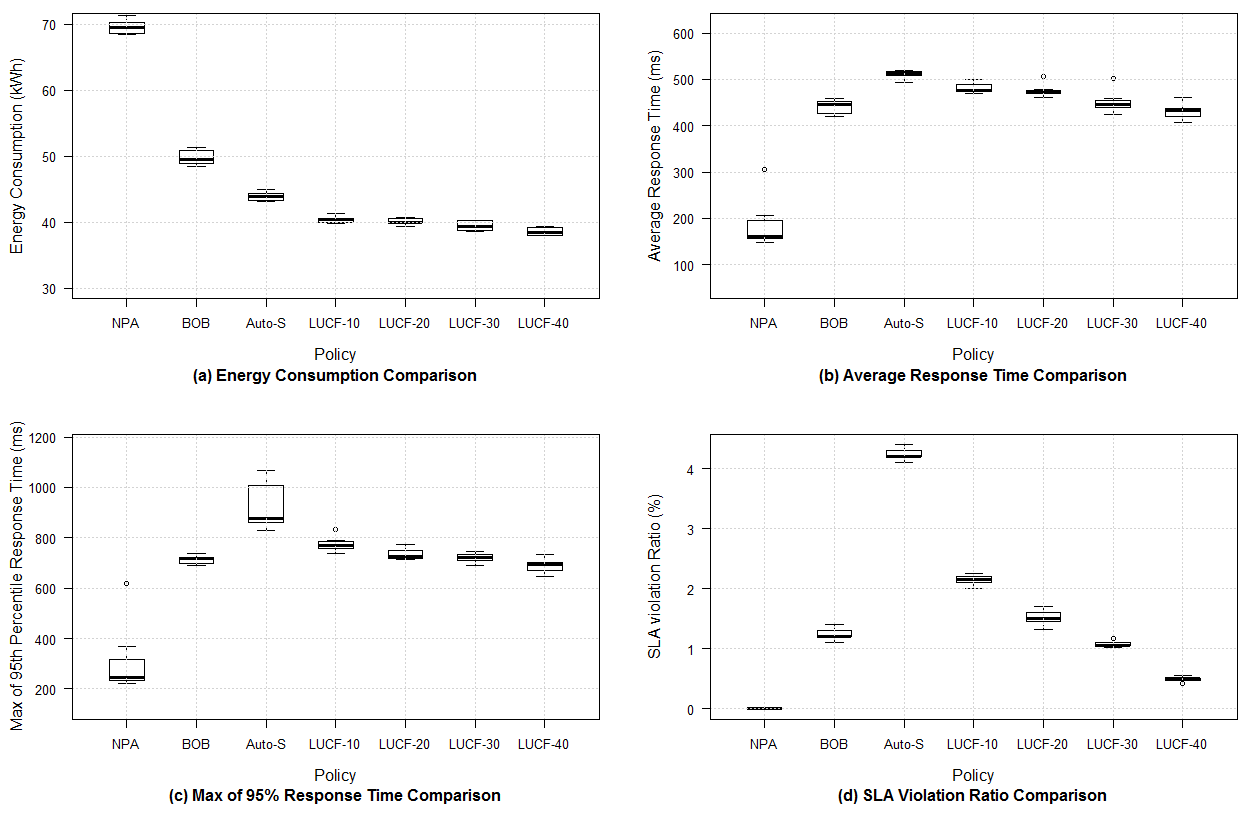}
	\caption[VarPerOptCom]{\color{black} Number of Active Hosts Comparison}
	
	\label{fig:VarPerOptCom}
\end{figure*}

\color{black} Fig. 6 shows the results with varied overloaded thresholds and optional utilization percentages for our proposed policies, we compare the energy consumption, average response time, maximum of 95th percentile response time and SLA violations achieved by LUCF, MNCF and RSC. For the energy consumption, under same optional utilization percentage, policies with 70\% overloaded threshold save more energy than policies with 80\%. For example, when the optional utilization percentage is 10\%, LUCF with 70\% overloaded threshold has 39.7 kWh and LUCF with 80\% overloaded threshold has 40.9 kWh. 
It is observed that with more optional utilization percentage, all the policies reduce more energy consumption, and both LUCF and MNCF save more energy consumption and RSC. Under 80\% overloaded threshold, as the energy consumption of LUCF and MNCF is quite close, we conduct the paired \textit{t}-tests for them, and the p-values are 0.09, 0.15, 0.1 and 0.09 respectively. Therefore, we conclude that energy consumption of LUCF and MNCF has no statistically significant difference when the overloaded threshold is 80\%. 

For the comparison of average response time and maximum of 95th percentile response time in Fig. 6(b) and Fig. 6(c), policies with 70\% overloaded threshold experience more average response time and maximum of 95th percentile response time than the ones with 80\% overloaded threshold. The average response time of LUCF with 70\% overloaded threshold ranges from 515 to 621 ms, while with 80\% overloaded threshold, it is from  452 ms to 500 ms. 
When more optional utilization percentage is configured, the average response time and the maximum of 95th percentile response time is reduced. For instance, with 80\% overloaded threshold, the average response time of LUCF is reduced from 500 to 452 ms, and the maximum of 95th percentile response time of MNCF is decreased from 780 to 680 ms. The results show that brownout-based policies are able to improve response time as well as energy saving. Fig. 6(d) illustrates the comparison of SLA violations. When the overloaded threshold is 70\% and optional utilization percentage is 10\% the SLA violation is more than 4\%, as the overloaded threshold and the optional utilization percentage increase, the SLA violations are reduced to less than 1\%. 

To conclude, LUCF and MNCF achieve better performance than RSC, as RSC selects containers randomly rather than deterministic methods. LUCF and MNCF have close energy consumption, but in most cases, LUCF achieves better performance in response time and SLA violations than MNCF. The reason lies in that LUCF has more container deactivation options than MNCF. For different overloaded thresholds comparison, policies with 70\% overloaded threshold save more energy but have the more average response time,  maximum of 95th percentile response time and SLA violations than policies with 80\% overloaded threshold. Configuring overloaded threshold as 80\% achieves a better trade-off than 70\%, as it reduces energy consumption while not having large average response time. Thus, the following experiments are conducted under 80\% overloaded threshold.  Additionally, as LUCF has the best performance among our proposed policies, we choose LUCF as the representative of our proposed algorithms to compare with benchmark policies.

  \color{black}

\begin{table*}[ht]
	\centering
	\caption{Final Experiment Results}
	\label{my-label}
	\begin{tabular}{ccccc}
		\textbf{Policy} & \textbf{Energy (kWh)} & \textbf{Average response time} & \textbf{Max of 95th response time} & \textbf{SLA violation} \\
		\hline \\
		NPA             & 69.71 (68.94,70.45)   & \textbf{188.8} (137.4, 240.2)           & \textbf{312.2} (178.8, 445.8)               & - \\
		\color{black}
		BOB             & \color{black} 49.83 (49.06, 50.60)   & \color{black} 440.1 (426.0, 454.1)           & \color{black} 712.4 (696.8, 727.9)                                    & \color{black} 1.240 (1.098, 1.381)                      \\	
			
		Auto-S          & 43.95 (43.48, 44.43)  & 511.0 (502.3, 519.6)           & 929.5 (840.9, 1018.1)              & 4.240 (4.098, 4.382)   \\
		LUCF-10         & 40.36 (40.01, 40.71)  & 482.1 (471.5, 492.7)           & 775.4 (746.2, 804.6)               & 2.140 (2.020, 2.259)   \\
		LUCF-20         & 40.17 (39.87, 40.47)  & 476.0 (462.4, 489.5)           & 735.7 (712.2, 759.1)               & 1.516 (1.340, 1.691)  \\
		LUCF-30         & 39.41 (38.93, 39.89)  & 451.5 (428.1, 475.0)           & 721.1 (702.3, 739.9)               & 1.082 (1.005, 1.158)  \\
		LUCF-40         & \textbf{38.60} (38.21, 39.01)  & 431.1 (415.0, 447.2)           & 687.8 (661.2, 714.4)               & \textbf{0.494} (0.439, 0.548) \\ 
		\hline
	\end{tabular}
\end{table*}

\begin{figure*}[!ht]
	\centering
	\includegraphics[width=0.85\linewidth]{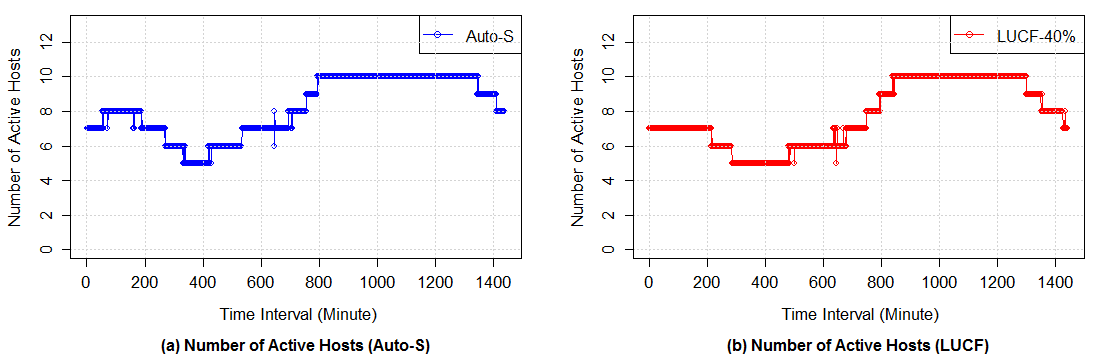}
	\caption[VarPerOptCom]{Number of Active Hosts Comparison}
	
	\label{fig:VarPerOptCom}
\end{figure*}

\begin{figure*}[!ht]
	\centering
	\includegraphics[width=1.0\linewidth]{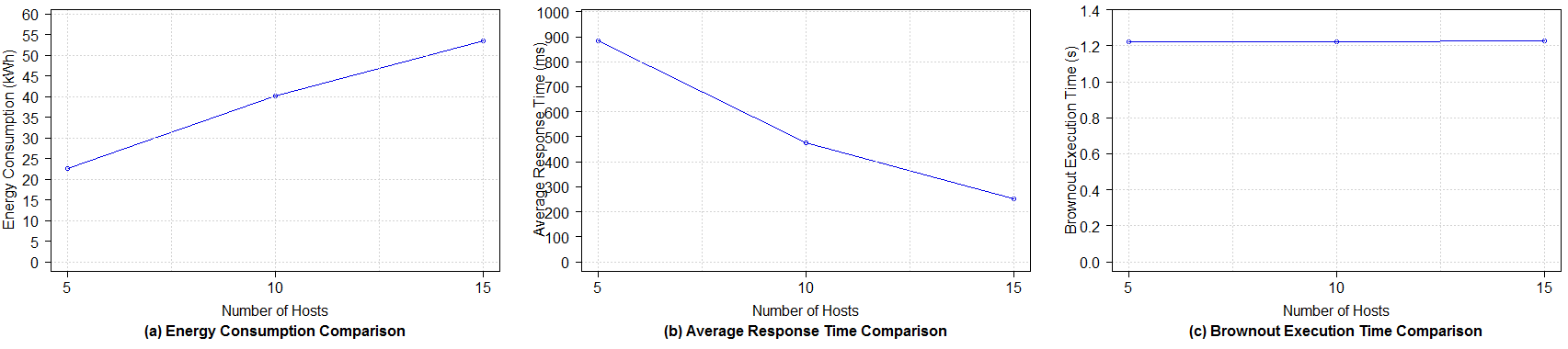}
	\caption[VarPerOptCom]{\color{black} Scalability Evaluation of iBrownout with LUCF policy}
	
	\label{fig:VarPerOptCom}
\end{figure*}

\begin{table}[!ht]
	\centering
	\caption{Scalability Experiments Results}
	\label{my-label}
					\scriptsize
	\begin{tabular}{|c|c|c|c|}
		\hline
		\textbf{Number of Hosts} & \textbf{\begin{tabular}[c]{@{}c@{}}Energy \\ Consumption\end{tabular}} & \textbf{\begin{tabular}[c]{@{}c@{}}Average \\ Response Time\end{tabular}} & \textbf{\begin{tabular}[c]{@{}c@{}}Brownout \\ Execution Time\end{tabular}} \\ \hline
		5 hosts                  & 22.6 kWh                                                               & 882 ms                                                                    & 1.223421 s                                                                  \\ \hline
		10 hosts                 & 40.2 kWh                                                               & 476 ms                                                                    & 1.224356 s                                                                  \\ \hline
		15 hosts                 & 53.4 kWh                                                               & 251 ms                                                                    & 1.224973 s                                                                  \\ \hline
	\end{tabular}
\end{table}

\textit{\textbf{(3) Final experiment results}}

\color{black}
Fig. 7 and Table 4 present the mean values of energy consumption, average response time, maximum of 95th percentile response time and SLA violations along with 95\% CI for the NPA, BOB, Auto-S and LUCF with different optional utilization percentages. The results demonstrate that NPA has energy consumption 69.71 kWh with 95\% CI (68.94, 70.45), BOB has 49.83 kWh with 95\% CI (49.06, 50.6), and Auto-S reduces it to 43.95 kWh with 95\% CI (43.48, 44.43). LUCF saves more energy consumption than Auto-S, to be more specific, LUCF with 10\% optional utilization leads to 40.36 kWh with 95\% CI (40.01, 40.71) and lowers gradually to 38.6 kWh with 95\% CI (38.21, 39.01) when optional utilization is 40\%. 

In the comparison of average response time and the maximum of 95th percentile response time in Fig. 7(b) and 7(c), as NPA has adequate resources,  it has the minimum response time compared with other policies.  Its average response time is 188.8 ms with 95\% CI (137.4, 240.2) and its maximum of 95th percentile response time is 312.2 ms with 95\% CI (178.8, 445.8).  
As Auto-S experiences overloads, its average response time and the maximum of 95th response time are 511 ms with 95\% CI (502.3, 519.6) and 929.5 with 95\% CI (840.9, 1018.1) respectively. 
Taking advantage of brownout, although BOB and LUCF endure overloads, their brownout controllers relieve the overloaded situation. In BOB, its average response time is reduced to 440.1 ms with 95\% CI (426.0, 454.1) and its maximum of 95th response time is 712.4 ms with 95\% CI (696.8, 727.9).
In LUCF with 40\% optional utilization percentage, its average response time and the maximum of 95th response time are reduced to 431.1 ms with 95\% CI (415, 447.2) and 687.8 ms with 95\% CI (661.2, 714.4) respectively. 
Fig. 7(d) presents the SLA violation comparison. NPA does not have SLA violations, BOB has 1.24\% with 95\% CI (1.098, 1.381), and Auto-S has 4.24\% with 95\% CI (4.098, 4.382) SLA violations.  When more optional utilization is offered, LUCF improves the SLA violations from 2.14\% to 0.5\% in average values. 

 This is due to the fact  that LUCF uses less active hosts as shown in Fig. 8, which shows the number of active hosts within one day. For instance, at the time intervals from 400-500, 6 hosts are active with Auto-S, while LUCF runs 5 active hosts.  For NPA and BOB, hosts are always at active states. From the presented results, we can conclude that the LUCF achieves better energy consumption than NPA, BOB and Auto-S. According to response time and SLA violation comparison, LUCF outperforms Auto-S. Compared with BOB, LUCF has better performance when optional utilization percentage is larger than 30\%.

\subsection{Scalability}
\color{black}
In this section, we evaluate the scalability of the proposed approach and the efficiency of the algorithm when the number of nodes is increased. As mentioned in previous sections, iBrownout is implemented based on Docker Swarm, thus, its performance depends on the performance of Docker Swarm. Our aim in this paper is not to discuss the scalability design of Docker Swarm. In \cite{Luzzardi}, the authors conducted scalability testing on Docker Swarm with 1,000 nodes and 30,000 containers, and results show that Docker Swarm has high scalability. 

We evaluate the scalability of iBrownout in terms of the number of hosts. The experiment settings are almost as same as in the previous experiments, the overloaded threshold is set as 80\% and optional utilization percentage is 30\%, while the difference lies in the number of hosts, we conduct experiments with 5, 10 and 15 hosts respectively.
Energy consumption and QoS are the main concern of our proposed approach. Because of page limitation, we only focus on average response time as QoS metric. In addition, to compare algorithm efficiency, we also evaluate the brownout algorithm (LUCF policy) execution time, which represents the time between brownout is triggered and the deactivated components are selected.

Fig. 9 and Table 5 show the impact of the varied number of hosts on energy consumption, average response time and brownout algorithm execution time. As it can be seen, when there are more hosts, the energy consumption is increased and the average response time is reduced, while the brownout execution time is kept as stable. The energy consumption is growing from 22.6 kWh with 5 hosts to 53.4 kWh with 15 hosts, while the average response time is dropping to 251 ms with 15 hosts from 882 ms with 5 hosts. The reason lies in that when more hosts are running, these hosts consume more energy, and the benefit is that the average response time is reduced due to more resources. 
The brownout execution time remains 1.22 s when the number of hots is varied. As mentioned in Section 6.3.1, although the algorithm complexity of LUCF is relevant to the number of hosts, the search operation in LUCF only consumes a small portion of time compared with the network delay to fetch the information of hosts and containers. Therefore, the brownout execution time remains stable when the number of hosts is increased. 
The results show that iBrownout scales reasonably well when the number of hosts grows. To be noted, the master node in Docker Swarm may be the bottleneck if there are a number of worker nodes but only one master node, thus, more nodes should be promoted as master nodes to ensure the system scalability.

\color{black}
\section{Conclusions and Future Work}
Brownout has been proven to be effective to solve the overloaded situation in cloud data centers. Additionally, brownout can also be applied to reduce energy consumption. 
In this paper, we introduced a brownout-based architecture by deactivating optional containers in applications or microservices temporarily to reduce energy consumption. Under this architecture, we introduce an integrated approach to managing energy and brownout in container-based clouds. We also propose several policies to find the suitable containers to deactivate and evaluate their performance in a prototype system. The experiment results under real test-beds have shown that our proposed policies achieve better performance in energy consumption, response time and SLA violations than baselines.

In the future, we plan to explore how brownout approaches can be applied in existing different approaches that are using models such as 1) Map-Reduce application 2) Stream-oriented application workload and 3) Bag of tasks application.


%

\appendices


\section*{Acknowledgments}
This work is supported by China Scholarship Council, Australia Research Council Future Fellowship and Discovery Project Grants. We thank Marcos Assuncao and Laurent Lefevre from INRIA (France) for providing the access to Grid'5000 infrastructure. \color{black} We also thank Shashikant Ilager for polishing the writing of this paper. \color{black}
Experiments presented in this paper were carried out using the Grid'5000 testbed, supported by a scientific interest group hosted by Inria and including CNRS, RENATER and several Universities as well as other organizations (see https://www.grid5000.fr).

\ifCLASSOPTIONcaptionsoff
  \newpage
\fi



%
%
%

\bibliographystyle{IEEEtran}
\bibliography{tsucsp}

%




\begin{IEEEbiography}
	[{\includegraphics[width=1in,height=1.25in,clip,keepaspectratio]{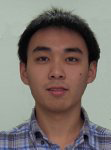}}]{Minxian Xu}
	received the BSc degree
	in 2012 and the MSc degree in 2015, both in
	software engineering
	from University of Electronic Science and Technology of China.
	He is working towards the PhD degree at the
	Cloud Computing and Distributed Systems
	(CLOUDS) Laboratory, School of Computing
	and Information Systems, the University of
	Melbourne, Australia. His research interests include resource scheduling and optimization in cloud computing. He has co-authored several peer-reviewed papers on T-SUSC, T-ASE, CCPE, ICSOC and ICC.
\end{IEEEbiography}
\begin{IEEEbiography}
	[{\includegraphics[width=1in,height=1.25in,clip,keepaspectratio]{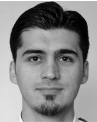}}]{Adel Nadjaran Toosi Dastjerdi}
 is a Research Fellow at the Cloud Computing and Distributed Systems (CLOUDS) Laboratory, School of Computing and Information Systems (CIS), University of Melbourne, Australia. He received his B.Sc. degree in 2003 and his M.Sc. degree in 2006 both in Computer Science and Software Engineering from Ferdowsi University of Mashhad, Iran and his Ph.D. degree in 2015 from the University of Melbourne.  Adel's Ph.D. studies were supported by International Research Scholarship (MIRS) and Melbourne International Fee Remission Scholarship (MIFRS). His Ph.D. thesis was nominated for CORE John Makepeace Bennett Award for the Australasian Distinguished Doctoral Dissertation and John Melvin Memorial Scholarship for the Best Ph.D. thesis in Engineering. His research interests include scheduling and resource provisioning mechanisms for distributed systems. Currently, he is working on resource management in Software-Defined Networks (SDN)-enabled Cloud Computing. 
	
\end{IEEEbiography}
\begin{IEEEbiography}
	[{\includegraphics[width=1in,height=1.25in,clip,keepaspectratio]{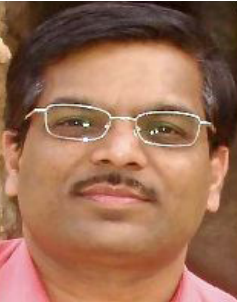}}]{Rajkumar Buyya}
 is a Redmond Barry Distinguished Professor and Director of the Cloud Computing and Distributed Systems (CLOUDS) Laboratory at the University of Melbourne, Australia. He is also serving as the founding CEO of Manjrasoft, a spin-off company of the University, commercializing its innovations in Cloud Computing. He served as a Future Fellow of the Australian Research Council during 2012-2016. He has authored over 625 publications and seven text books including "Mastering Cloud Computing"  published by McGraw Hill, China Machine Press, and Morgan Kaufmann for  Indian, Chinese and international markets respectively.  He is one of the highly cited authors in computer
science and software engineering worldwide (h-index=114, g-index=245, 66,900+ citations).  Dr. Buyya is recognized as a "Web of Science Highly Cited Researcher" in 2016 and 2017 by Thomson Reuters, a Fellow of IEEE, and Scopus Researcher of the Year 2017 with Excellence in Innovative Research Award by Elsevier for his outstanding contributions to Cloud computing. He served as the founding Editor-in-Chief of the IEEE Transactions on Cloud Computing. He is currently serving as Co-Editor-in-Chief of Journal of Software: Practice and Experience, which was established over 45 years ago.
For further information on Dr.Buyya, please visit his cyberhome:
www.buyya.com
	
\end{IEEEbiography}

\end{document}